\documentclass[showpacs,preprintnumbers,amsmath,amssymb]{revtex4}

\usepackage{graphicx}
\usepackage{bm}

\DeclareMathOperator{\sign}{sign}

\begin{document}

\title{Existence of breathing patterns in globally coupled finite-size
nonlinear lattices}

\author{Dirk Hennig}
\affiliation{Department of Mathematics, University of Portsmouth, Portsmouth, PO1 3HF, UK}

\date{\today}

\begin{abstract}
\noindent 
We prove the existence of  time-periodic solutions consisting of patterns 
built up from two states, one with small amplitude and the other one with large amplitude, 
in general  nonlinear Hamiltonian  finite-size lattices with global coupling.
Utilising 
the comparison
principle for differential equations it is  
demonstrated that for a two site segment of the nonlinear lattice 
one can construct  solutions  that are 
sandwiched between two oscillatory localised lattice states.  
Subsequently, it is proven that such a localised state can be embedded in the 
extended nonlinear lattice forming a breathing pattern   
with a single site of large amplitude against a background of uniform small-amplitude 
states. Furthermore, it is demonstrated that spatial patterns are possible that  
are built up from any combination of the small-amplitude state and 
the large-amplitude state.
It is shown that for soft (hard) on-site potentials  
the range of allowed frequencies of the in-phase (out-of-phase) breathing patterns
extends to values   
 below (above) the lower (upper) value of the bivalued degenerate linear spectrum of phonon frequencies.
\end{abstract}

\pacs{05.45.-a, 63.20.Pw, 45.05.+x, 63.20.Ry}
\maketitle

\noindent Intrinsic localised modes (ILMs) or discrete breathers in nonlinear lattices
have  attracted significant interest recently, not least due to the important role they play in many 
physical realms where features 
of localisation in systems of coupled oscillators are involved 
(for a review see \cite{Flach1} and references
therein),\cite{Aubry}-\cite{PhysicsReports}. 
Proofs of existence and nonexistence
of breathers, as spatially localised and time-periodically varying solutions,   were 
provided in \cite{MacKay}-\cite{Dirk}. The exponential stability of breathers 
was proven in \cite{Bambusi}. Analytical and numerical methods have been
developed to continue breather solutions in conservative and dissipative systems  
starting from the anti-integrable limit \cite{Sepulchere}-\cite{Martinez2}. 

During recent years the
existence of breathers has been verified in a number 
of experiments in various contexts including micro-mechanical cantilever arrays \cite{cantilever}, 
arrays of coupled Josephson junctions \cite{josephson}, antiferromagnetic chains \cite{antiferro}, coupled optical wave guides \cite{optical}, 
Bose-Einstein condensates in optical lattices \cite{BEC}, in coupled torsion pendula \cite{Jesus}, 
electrical transmission lines \cite{electrical}, and granular crystals \cite{crystals}. 
Regarding their creation mechanism in conservative systems, modulational instability (MI) provides the route to 
the formation of breathers originating from an initially spatially homogeneous state imposed to (weak) perturbations. 
To be precise, the MI of band edge plane waves 
triggers an inherent instability leading to the formation of a spatially  localised state 
\cite{Remoissenet}. 

Lattice discreteness and nonlinearity are the prerequisites for 
the ability of breathers to store energy without dispersing it to the environs.
Discreteness serves for the existence of an upper boundary to the linear phonon spectrum making
it possible that not only the fundamental 
breather frequency but also all of its higher harmonics lie outside of the phonon spectrum. Nonlinearity 
allows to tune the frequency 
of (anharmonic) oscillators in dependence of their amplitude (or action) such that resonances between harmonics
of the breather frequency, $\omega_b$, and the 
frequencies, $\omega_{ph}$, of linear oscillations, i.e. phonons, can be avoided. That is, the non-resonance conditions 
\begin{equation}
 m \omega_b \ne \omega_{ph},\,\,\,\forall m \in {\mathbb{Z}}
\end{equation}
need to be satisfied.
 In finite  systems the separation between phonon 
frequencies allows that a breather frequency lies in a non-resonance window between  two frequencies in the phonon 
spectrum \cite{Marin},\cite{Morgante}.   In contrast, this is impossible for localised solutions in infinite systems 
as the phonon spectrum is characterised by continuous bands.

In this work we prove the existence of  time-periodic solutions consisting of patterns 
built up from 
two states of different amplitude (one  small and the other one large)
in general  nonlinear finite-size lattices with global coupling. We treat 
two types of on-site potentials;
namely hard and soft ones.
Unlike for the continuation process of (trivially) localised solutions 
starting from the anti-continuum limit  \cite{Marin},\cite{Aubry1}-\cite{Martinez2} 
our approach is not necessarily confined to the weak coupling regime.

We study the dynamics of general  nonlinear lattice systems
given by the following equations
\begin{eqnarray}
\frac{d^2q_n(t)}{dt^2}&=&-U^{\prime}(q_n(t))+\frac{\kappa}{2N+1} \sum_{j =-N}^{N} (q_{n+j}(t)-q_n(t)),
\label{eq:start}
\end{eqnarray}
with $-N\le n  \le N$, $q_{n>N}=q_{n-(2N+1)}$ and $q_{n<-N}=q_{n+(2N+1)}$ 
and the prime $^{\prime}$ stands for the derivative with respect to $q_n$, the latter being 
the coordinate of the oscillator at site $n$ evolving in an anharmonic on-site potential 
$U(q_n)$.
Each oscillator interacts with the remaining oscillators amounting to 
global diffusive coupling, the strength of 
which is regulated by the value of the parameter $\kappa$.
Even though the system in its formulation in (\ref{eq:start}) is associated with a lattice having an odd 
number of sites, $2N+1$, we emphasise that our proofs of the 
existence of breathing pattern solutions cover also systems 
with an even number of sites $2N$ (see below).

The on-site potential $U$ is analytic  and is assumed to have the following properties:
\begin{equation}
U(0)=U^{\prime}(0)=0\,,\,\,\,U^{\prime \prime}(0)> 0.\label{eq:assumptions}
\end{equation}
In what follows we differentiate between soft on-site potentials and hard on-site potentials. For the former (latter)
the oscillation frequency of an oscillator moving in the on-site potential $U(q)$
decreases (increases) with increasing oscillation amplitude.
A soft potential possesses at least one inflection point. If a soft potential possesses a 
single inflection point, denoted by $q_i$, 
we suppose without loss of generality  (w.l.o.g.) that $q_i>0$. Then the following relations are valid
\begin{eqnarray}
U^{\prime}(-\infty <q<0)&<&0,\,\,\, U^{\prime}(0<q<q_i)>0,\\
U^{\prime \prime}(q_{i})&=&0,\,\,\,  U^{\prime \prime }(-\infty<q<q_i)>0.
\end{eqnarray} 
If $U(q)$ possesses two inflection points denoted by $q_{i,-}<0$ and $q_{i,+}>0$  it holds that 
\begin{eqnarray}
U^{\prime}(q_{i,-}<q<0)&<&0,\,\,\,  U^{\prime}(0<q<q_{i,+})>0\label{eq:U1prime}\\
U^{\prime \prime}(q_{i,\pm})&=&0,\,\,\,  U^{\prime \prime}(q_{i,-}<q<q_{i,+})>0.\label{eq:U2prime}
\end{eqnarray}
We remark that $U(q)$ can have more than two inflection points (an example is a periodic potential $U(q)=-\cos(q)$). 
However, in the frame of the current study we are only interested in motion 
between the inflection points adjacent to the minimum of $U(q)$ at $q=0$.  

Hard on-site potentials are, in addition to the assumptions in (\ref{eq:assumptions}), 
characterised in their entire range of definition  by 
\begin{equation}
U^{\prime}(q<0)<0,\,\,\,U^{\prime}(q>0)>0,\,\,\,U^{\prime \prime}(q)> 0.\label{eq:assumptionhard} 
\end{equation}

The system of equations (\ref{eq:start}) possesses an energy integral 
\begin{equation}
 E=\sum_{n=-N}^N\left[\frac{1}{2}\dot{q}_n^2+U(q_n) \right]+
 \frac{\kappa}{2(2N+1)} \sum_{n=-N}^{N-1}\sum_{m=n+1}^{N}(q_{n}-q_m)^2.
\end{equation}
There exists a closed maximum equipotential surface 
$\sum_{n=-N}^N U(q_n)+\kappa/(2(2N+1)) \sum_{n=-N}^{N-1}\sum_{m=n+1}^{N}(q_{n}-q_m)^2=E$ bounding all motions and one has
$\dot{q}_{-N}=\,...\,=\dot{q}_N=0$ on this surface.
   
The linearised system exhibits plane wave (phonon) solutions with frequencies 
\begin{equation}
 \omega_{ph}^2(k)= \omega_{0}^2+\frac{4\kappa}
 {2N+1}\sum_{n=1}^N \sin^2\left(\frac{n k}{2}\right),\,\,\,k=\frac{2\pi m}{2N+1},
 \,\,\,m=0,\pm 1,...,\pm N,\label{eq:linear}
\end{equation}
and $\omega_0^2=U^{\prime \prime}(0)$.
Since
\begin{equation}
\frac{4}{2N+1}\sum_{n=1}^N \sin^2\left(\frac{n \pi m}{2N+1}\right) =1
\end{equation}
for $m=\pm 1,...,\pm N$, the linear spectrum consists of two frequency values only; the {\it lower} value
\begin{equation}
\omega_{ph}^2(k=0)=\omega_0^2,
\end{equation}
and the  $2N$-fold degenerate  {\it upper} value 
\begin{equation}
\omega_{ph}^2(k\ne 0)=\omega_{0}^2+\kappa.
\end{equation}

We study the existence of solutions to the system (\ref{eq:start}) that are 
 time-periodic satisfying
 \begin{equation}
  q_n(t+T_b)=q_n(t)
 \end{equation}
with period $T_b=2\pi/\omega_b$. Moreover, these periodic solutions are supposed comprise 
two different states, one with a small amplitude and the other one with a large amplitude.

\vspace*{1.0cm}

The forthcoming derivations of estimates are facilitated by the following statement:

\vspace*{0.5cm}

\noindent {\bf Lemma:}  For  soft potentials $U(q)$ with two
inflection points $q_{i,\pm}$ consider the set
\begin{equation}
I_s:=\left\{q\,|\, q_{i,-}<q_{l}\le q \le  q_r<q_{i,+}\right\}.
\end{equation}

Then it holds that  for any pair $x,y \in I_s$ with $x< y$ 
\begin{equation}
U^{\prime \prime}(0)\left(y-x\right)>\left[U^{\prime}(y)-U^{\prime}(x)\right] > 
\min\{U^{\prime \prime}(q_{l}),U^{\prime \prime}(q_{r})\}\left(y-x\right)>0.\label{eq:deltabounds}
\end{equation}

For hard potentials consider the set 
\begin{equation}
I_h:=\left\{q\,|\, -\infty<Q_l \le q \le Q_r<\infty \right\}.
\end{equation}
Then it holds that  for any pair $x,y \in I_h$ with $x<y$
\begin{equation}
\max \{U^{\prime \prime}(Q_l),U^{\prime \prime}(Q_r)\}\left(y-x\right)>\left[U^{\prime}(y)-U^{\prime}(x) \right] > U^{\prime \prime}(0) 
(y-x)>0.\label{eq:deltaboundh}
\end{equation}

\vspace*{0.5cm}

\noindent {\bf Proof:} Consider the expression
\begin{equation}
F(x,y)=\frac{ U^{\prime}(y)-U^{\prime}(x) }{y-x}.
\end{equation}
Note that 
\begin{equation}
 \lim_{y \rightarrow x} \frac{ U^{\prime}(y)-U^{\prime}(x) }{y-x}=U^{\prime \prime}(x) > 0.
\end{equation}

By assumptions (\ref{eq:U2prime}) and (\ref{eq:assumptionhard}) we have that on the sets 
$I_s$ and $I_h$ it holds that  
$U^{\prime}(y)>U^{\prime}(x)$ for $y>x$. Therefore the expression $F(x,y)$ is positive.
Furthermore, by virtue of the mean value theorem there exist a point $z$ in $(x,y)$ such that 
\begin{equation}
\frac{U^{\prime}(y)-U^{\prime}(x) }{y-x}=U^{\prime \prime}(z) \ge 
\min_{q \in I_s,I_h} U^{\prime \prime}(q),
\end{equation}
and 
\begin{equation}
 \frac{U^{\prime}(y)-U^{\prime}(x)}{y-x}=U^{\prime \prime}(z)\le 
 \max_{q \in I_s,I_h}\left\{U^{\prime \prime}(q)\right\}.
\end{equation}

One has for  soft potentials  $\min_{q \in I_s}( U^{\prime \prime}(q))=
\min\{U^{\prime \prime}(q_{l}),U^{\prime \prime}(q_{r})\}$,
and 
$\max_{q \in I_s}( U^{\prime \prime}(q))=U^{\prime \prime}(0)$ therefore 
it holds that 
\begin{equation}
 U^{\prime \prime}(0)(y-x)\ge U^{\prime}(y)-U^{\prime}(x)\ge 
 \min\{U^{\prime \prime}(q_{l}),U^{\prime \prime}(q_{r})\}(y-x)>0.
\end{equation}

Similarly for  hard potentials by the assumption (\ref{eq:assumptions}) one has 
$\min_{q \in I_h} (U^{\prime \prime}(q))=U^{\prime \prime}(0)$ and 
$\max_{q \in I_h} (U^{\prime \prime}(q))=
\max\{U^{\prime \prime}(Q_l),U^{\prime \prime}(Q_r)\}$,
so that \begin{equation}
\max\{U^{\prime \prime}(Q_l),U^{\prime \prime}(Q_r)\}(y-x)\ge U^{\prime}(y)-U^{\prime}(x)\ge U^{\prime \prime}(0)(y-x)>0\nonumber
\end{equation} 
completing the proof.

\hspace{16.5cm} $\square$

\vspace*{0.5cm}
Remark: To apply the Lemma  to the case of  
soft potentials with a single inflection point $q_i>0$ one proceeds along the lines given above 
for the Lemma considering the set $\{q\,|\,-\infty <q_l \le q \le q_r<q_i\}$. 

\vspace*{1.0cm}
As a first step towards the proof of existence of breather solutions to 
the system (\ref{eq:start}) we treat the reduced system of two 
interacting oscillators.

\section{Localised solutions for two coupled nonlinear oscillators}\label{section:two}

In this section we study the dynamics of two interacting general nonlinear oscillators  given by the 
following system
\begin{eqnarray}
\ddot{q}_1(t)&=&-U^{\prime}(q_1(t))+\kappa (q_2(t)-q_1(t))
\label{eq:start1}\\
\ddot{q}_2(t)&=&-U^{\prime}(q_2(t))-\kappa (q_2(t)-q_1(t)).
\label{eq:start2}
\end{eqnarray}
As above the prime $^{\prime}$ stands for the derivative with respect to $q_n$ and an overdot $\dot{}$ 
 represents the derivative with respect to time $t$. 

In the following we prove the existence of spatially localised and time-periodic solutions 
for the system\,(\ref{eq:start1}),(\ref{eq:start2}).  
The localised solutions, like  normal modes, are understood as a 
{\it vibration in unison} of the system, i.e. 
all units of the system perform synchronous oscillations 
so that the two oscillators pass through their extreme values simultaneously.

\vspace*{0.5cm}
\noindent {\bf Theorem 1:} {\it Let $(q_n(t),\dot{q}_n(t))$, $n=1,2$, be the smooth solutions 
to Eqs.\,(\ref{eq:start1}),(\ref{eq:start2}) with a soft on-site potential and
$q_{i,-}<q_{l}\le q_{1,2}(t) \le  q_r<q_{i,+}$ for $t \in {\mathbb{R}}$. 
There exist periodic solutions $(q_n(t+T_b),\dot{q}_n(t+T_b))=(q_n(t),\dot{q}_n(t))$ 
for  $n=1,2$,
so that   
the oscillators perform in-phase motion, i.e. 
$\sign (q_1(t))=\sign (q_{2}(t))$ with period $T_b=2\pi/\omega_b$ and frequencies
$\omega_b$ satisfying
\begin{equation}
\sqrt{\min \{U^{\prime \prime}(q_l),U^{\prime \prime}(q_r)\}}\le \omega_b < 
\sqrt{U^{\prime \prime}(0)+2\kappa}.
\label{eq:ineq1}
\end{equation}

Moreover, these  solutions are localised which is characterised by either 
\begin{equation}
 |q_1(t)|> |q_{2}(t)|,\,\,\,t\in {\mathbb{R}},\,\,\,t \neq {\tilde{t}_k},\,\,\,k \in {\mathbb{Z}}
\end{equation}
or 
\begin{equation}
 |q_2(t)|> |q_{1}(t)|,\,\,\,t\in {\mathbb{R}},\,\,\,t 
 \neq {\tilde{t}_k},\,\,\,k \in {\mathbb{Z}}
\end{equation}
where $\tilde{t}_{k \in {\mathbb{Z}}}$ are the instants of time when the
 oscillators pass simultaneously through zero coordinate,
corresponding to the minimum position of the on-site potential, i.e. 
$q_{1}(\tilde{t}_k)=q_{2}(\tilde{t}_k)=0$.
}

\vspace*{0.5cm}
\noindent {\bf Proof:} W.l.o.g. the initial conditions satisfy 
\begin{equation}
 q_{1}(0)=q_2(0) =0,\,\,\,  0<\dot{q}_{2}(0) <\dot{q}_{1}(0).\label{eq:ic2}
\end{equation}

Then due to continuity there must exist some 
$t_*>0$ so that during the interval $(0,t_*]$ the 
following  order relation is satisfied
\begin{equation}
 0 < q_{2}(t) < q_{1}(t).\label{eq:order1}
\end{equation}

We define the difference variable between the coordinates at sites $n=1$ and $n=2$ as follows
\begin{equation}
\Delta q_*(t)=  q_{1}(t)-  q_{2}(t).
\end{equation} 
Thus by definition  $\Delta q_*(t) \ge 0$ on $[0,t_*]$.

The time evolution of the difference variables  $\Delta q_* (t)$ is 
determined by the following equation 
\begin{equation}
 \frac{d^2 \Delta q_*}{dt^2}= -\left[U^{\prime}(q_{1})-U^{\prime}(q_{2})\right]-
 2\kappa \Delta q_*.
  \label{eq:deltap}
\end{equation}

Discarding the non-positive term $-2{\kappa} \Delta q_*$  for $q_1(t) \ge q_2(t)$ and utilising the Lemma above, 
we bound the r.h.s. of Eq.\,(\ref{eq:deltap})  
from above as follows: 
\begin{equation}
\frac{d^2 \Delta q_*}{dt^2}\le -\Omega^2_s\Delta q_*(t)
\end{equation}
where $\Omega_s^2=\min\{U^{\prime \prime}(q_l),U^{\prime \prime}(q_r)\}$.

Similarly we bound the r.h.s. of Eq.\,(\ref{eq:deltap})
for $q_1(t) \ge q_2(t)$
from below as follows:
\begin{equation}
 \frac{d^2 \Delta q_*}{dt^2}\ge -(\omega^2_0+2{\kappa})\Delta q_*(t),
\end{equation}
with $\omega^2_0=U^{\prime \prime}(0)$. 

Therefore, by the comparison principle for differential equations, 
$\Delta q_*(t)$ and $\Delta \dot{q}_*(t)$  are bounded from above and below
for given initial conditions by the solution of 
\begin{equation}
\frac{d^2 a}{dt^2}=-\Omega^2_s  a
\label{eq:boundabove}
\end{equation}
and 
\begin{equation}
\frac{d^2 b}{dt^2}=-(\omega^2_0 +2{\kappa}) b,
\label{eq:boundbelow}
\end{equation}
respectively, provided $a(t)\ge 0$ and $b(t)\ge 0$.

The solution to Eq.\,(\ref{eq:boundabove}) and (\ref{eq:boundbelow}) with initial conditions 
$(\Delta q_*(0)=0,\Delta \dot{q}_*(0^+)\equiv \Delta \dot{q}_0 >  0)$ is given 
by 
\begin{eqnarray}
  a(t)&=& \frac{\Delta \dot{q}_0}{\Omega_s}\sin(\Omega_s t),\\ \label{eq:qabove}
  \dot{a}(t)&=& \Delta \dot{q}_0\cos(\Omega_s t) 
  \label{eq:pabove}
\end{eqnarray}
and 
\begin{eqnarray}
  b(t)&=& \frac{\Delta \dot{q}_0}{\sqrt{\omega^2_0+2{\kappa}}}
  \sin(\sqrt{\omega^2_0+2{\kappa}}\, t),\\ \label{eq:qbelw}
  \dot{b}(t)&=& \Delta \dot{q}_0\cos(\sqrt{\omega^2_0+2{\kappa}}\, t),
  \label{eq:pbelow}
\end{eqnarray}
respectively where
$a(t)\ge 0$ for $0 \le t \le  \pi/\Omega_s$ and $b(t)\ge 0$ 
for $0 \le t \le  \pi/\sqrt{\omega^2_0+2{\kappa}}$.
By $f(\tau_k^{-})$ and  $f(\tau_k^{+})$ the left-sided 
 and right-sided limits of $f(t)$ for $t \rightarrow \tau_k$ are 
 meant, respectively. 
 
Notice that $d^2 \Delta q_*(t)/dt^2< 0$ on $(0,t_*)$, that is, 
the acceleration stays negative.
Due to the relations $\Delta \dot{q}_0>0$  in conjunction with the lower bound
   $b(t) \le \Delta q_*(t)$  the order relation
as given in (\ref{eq:order1})   is at least maintained on the interval 
$(0,\pi/\sqrt{\omega^2_0+2{\kappa}})$. Moreover,  
$\Delta q_*(t)$ 
is bound to grow monotonically 
at least during the interval 
$(0,\pi/(2\sqrt{\omega^2_0+2{\kappa}})]$ 
and attains a least maximal value $\Delta \dot{q}_0/\sqrt{\omega^2_0+2{\kappa}}$. 
Furthermore, $\Delta q_*(t)$ cannot return to zero before 
$t=\pi/\sqrt{\omega^2_0+2{\kappa}}$.

From the upper bound $ \Delta q_*(t) \le a(t)$  one infers that $\Delta q_*(t)$
can attain an absolute maximal value 
$\Delta \dot{q}_0/\Omega_s$ but not before $t=\pi/(2\Omega_s)$ and 
$\Delta q_*(t)$ is bound
to return to zero not later than $t=\pi/\Omega_s$. 
Similarly, $\Delta \dot{q}_*(t)$ is bound to decrease monotonically for $0 <t \le  \pi/\Omega_s$ and 
becomes negative at a time in the interval\\  $(\pi/(2\sqrt{\omega^2_0+2{\kappa}}),\pi/(2\Omega_s))$. 
Moreover, it holds that $\Delta \dot{q}_*(t)\ge -\Delta \dot{q}_0$ for  
$0\le t\le \pi/\sqrt{\omega^2_0+2{\kappa}}$ and $\Delta \dot{q}_*(t)\le -\Delta \dot{q}_0$ 
for  $t\ge \pi/\Omega_s$.

Therefore, by the smooth dependence  of the solutions 
$(\Delta q(t), \Delta \dot{q}(t))$ on the initial values 
 $(q_{1,2}(0),\dot{q}_{1,2}(0))$, to any chosen initial 
condition $q_1(0)=0,\dot{q}_1(0)\ne 0$ and $q_2(0)=0$ there exists a corresponding $\dot{q}_2(0)$ 
(or vice versa) so that one has $\Delta q_*(t_{*})=\Delta q_*(0)=0$ and 
$\Delta \dot{q}_*(t_{*}^-)=-\Delta \dot{q}_0$ with
$t_* \in [\pi/\sqrt{\omega^2_0+2{\kappa}},\pi/\Omega_s]$.
This implies the symmetry  
\begin{eqnarray}
\Delta q_*(t_{*}/2+\tau)&=&\Delta q_*(t_{*}/2-\tau),\label{eq:sym1}\\
-\Delta \dot{q}_*(t_{*}/2+\tau))&=&\Delta \dot{q}_*(t_{*}/2-\tau))\label{eq:sym2}
\end{eqnarray}
with $0<\tau<t_{*}/2$ and $t_{*}/2$ corresponds to the turning point of the motion when
$\Delta q_*$   attain its maximum 
while $\Delta \dot{q}_*$  is zero. In turn, this implies that  the
motion of the two oscillators possesses the symmetry  
\begin{eqnarray}\
q_n(t_{*}/2+\tau)&=&q_n(t_{*}/2-\tau),\,\,\,n=1,2\\ 
\dot{q}_n(t_{*}/2+\tau)&=&\dot{q}_n(t_{*}/2-\tau)),\,\,\,n=1,2,
\end{eqnarray}
with $0\le\tau\le t_{*}/2$ and at the turning point, $t=t_{*}/2$,
the two coordinates
$q_{1}$ and $q_2$ 
assume simultaneously their 
respective maxima 
while $\dot{q}_{1}$ and $\dot{q}_2$ are both zero.
Conclusively, on the interval $[0,t_*]$ 
the two  oscillators evolve through half a  cycle of periodic 
in-phase motion, i.e. $\sign (q_1)=\sign (q_{2})$ and
$\sign(\dot{q}_1)=\sign(\dot{q}_{2})$.
With regard to later use we remark that the value of the coordinates $q_{1,2}$ 
at the turning point $t=t_*/2$ together with $\dot{q}_{1,2}=0$ can be taken as (new) initial conditions 
to obtain time-reversible solutions.

At  $t=t_*$, just as at $t=0$,  the  oscillators pass simultaneously through zero coordinate,
corresponding to the minimum position of the on-site potential at $q_n=0$ and the oscillators 
proceed afterwards with negative coordinate, i.e. $q_n(t)<0$ and 
negative velocity, i.e. $\dot{q}_n(t)<0$, $n=1,2$, 
until the next turning point is reached. 

Since at $t=t_*$ one has $\dot{q}_{1}(t_*)<\dot{q}_{2}(t_*)<0$ 
then due to continuity there must exist some 
$t_{**}>0$ so that during the interval $(t_*,t_{**}]$  
the  order relation $q_{1}(t) < q_{2}(t)$ is satisfied.

For $t\ge  t_*$ we consider the
difference variable between the coordinates at sites $n=1$ and $n=2$ 
as follows
\begin{equation}
\Delta q_{**}(t)=  q_{2}(t)-  q_{1}(t).
\end{equation}

Initialising the dynamics accordingly with 
$\Delta q_{**}(t_*)=\Delta q_{*}=0$, 
$\Delta \dot{q}_{**}(t_*^+)=-\Delta \dot{q}_{*}(t_*^-)=\Delta \dot{q}_*(0^+)$ 
and with the arguments given above it follows that $\Delta q_{**}(t)$ and $\Delta \dot{q}_{**}(t)$  exhibit 
 qualitatively the same features as $\Delta q_{*}(t)$ and $\Delta \dot{q}_{*}(t)$
 for $0\le t\le t_*$ so that lower and upper bounds on 
$\Delta q_{**}(t)$ and $\Delta \dot{q}_{**}(t)$ are derived equivalently to the ones above. 
In fact, one has 
\begin{eqnarray}
q_n((t_{*}+t_{**})/2+\tau)&=&q_n((t_*+t_{**})/2-\tau),\\
\dot{q}_n((t_*+t_{**})/2+\tau)&=&\dot{q}_n((t_*+t_{**})/2-\tau),
\end{eqnarray}
with $n=1,2$ and  $0 \le \tau\le (t_{**}-t_{*})/2$ and $(t_*+t_{**})/2$ 
corresponds to the turning point of the motion when the coordinates $q_{1,2}$  attain their 
minima 
while the velocities $\dot{q}_{1,2}$ become zero.
Again we remark with regard to later use that the value of the coordinates $q_{1,2}$ 
at the turning point $(t_*+t_{**})/2$ together with $\dot{q}_{1,2}=0$ can be taken as 
(new) initial conditions 
to generate time-reversible solutions.

In particular the time 
 $t_{**}$, at which 
 $\Delta q_{**}(t_{**})=0$  
 and $\Delta \dot{q}_{**}(t_{**}^-)=-\Delta \dot{q}_{**}(t_*^+)$, lies 
 in the range $t_*+\pi/\sqrt{\omega^2_0+2 \kappa} < t_{**}< t_*+\pi/\Omega_s$.   
 The zero of $\Delta q_{**}$ marks the end of a (first) cycle  of duration 
 $2\pi/\sqrt{\omega^2_0+2 \kappa} <T_b=t_{**}<2\pi/\Omega_s$ of 
 maintained localised oscillation throughout 
 of which the order relation $|q_1(t)|\ge |q_{2}(t)|$ is preserved. 
 
Notice that $t_*$ does not necessarily equals  $t_{**}-t_*$ 
when the oscillators perform motion in on-site potentials without 
reflection symmetry, viz. $U(q) \neq U(-q)$.
As the breather frequency $\omega_b=2\pi/T_b$ 
depends on the amplitude of oscillations, the latter can be chosen 
such that the non-resonance condition 
$m \omega_b \ne \omega_0$ 
for all $m \in \mathbb{Z}$ is satisfied. 

\vspace{0.5cm}

In relation to the time-periodicity of the dynamics of localised 
 solutions beyond times $t\ge t_{**}$ 
we  consider the dynamics of $\Delta q_k(t)=(-1)^{k+1}(q_1(t)-q_2(t))$ on intervals 
 \begin{equation}
  I_k:=[t_k,t_{k+1}],\,\,\,\mbox{with integer}\,\,\, k\ge 1,\,\,\,t_1=t_{**}\label{eq:interval1}
\end{equation}
 with  
 \begin{equation}
  t_{k+1} =\left\{ \begin{array}{cl} t_k+t_* & \mbox{for}\,\,\,k\,\,\,  \mbox{odd} \\
                               t_k+ t_{**}-t_* & \mbox{for}\,\,\,k\,\,\, \mbox{even}\,. \\
                              \end{array} \right.\label{eq:interval2}
 \end{equation}
Crucially,  initialising the dynamics with 
$\Delta q_k(t_k)=0$ and 
$\Delta \dot{q}_k(t_k^+)=\Delta q_0$
on each of 
the intervals $I_k$, $\Delta q(t)$ and $\Delta \dot{q}(t)$   
periodically repeat the behaviour 
described above for the interval $[0,t_*]$ 
for odd $k$ and, $[t_*,t_{**}]$ for even $k$. 
Conclusively, 
spatially localised and 
time-periodic  solutions result satisfying
\begin{equation}
 |q_1(t)|\ge |q_{2}(t)|,\label{eq:final}
\end{equation}
and $(q_n(t+T_b),\dot{q}_n(t+T_b))=(q_n(t),\dot{q}_n(t))$ for $n=1,2$ with 
period $T_b=2\pi/\omega_b$ 
where the frequency $\omega_b$ satisfies the relation
\begin{equation}
 \sqrt{\min\{U^{\prime \prime}(q_l),U^{\prime \prime}(q_r)\}}\le \omega_b
 < \sqrt{U^{\prime \prime}(0)+2\kappa}.\label{eq:ranges1}
\end{equation}

For later use we denote the small (large)
amplitude state by $\{s\}$ ($\{L\}$).
Note that the inequality in (\ref{eq:final}) is  strict except  at those moments in time,  
$\tilde{t}_{k\in {\mathbb{Z}}}$, when the
 oscillators pass simultaneously through zero coordinate,
corresponding to the minimum position of the on-site potential, i.e. 
$q_{1,2}(\tilde{t}_k)=0$.

From the discussion above we conclude that 
there exist initial conditions such that the associated periodic solutions possess 
the time-reversibility symmetry, viz. $q_{1,2}(t)=q_{1,2}(-t)$ 
and $-\dot{q}_{1,2}(t)=\dot{q}_{1,2}(-t)$ so that 
the relation (\ref{eq:final}) is true for $t \in {\mathbb{R}}$. 
In particular, by taking these  initial conditions as 
$q_{i,-}<q_{1,2}(0)<q_{i,+}$, and $q_{1,2}(0) \ne 0$,  
$\dot{q}_{1,2}(0)=0$, the relations $q_{i,-}<q_{l}\le q_{1,2}(t) \le  q_r<q_{i,+}$ and 
either $|q_2(t)|\le |q_1(t)|$ or $|q_1(t)|\le |q_2(t)|$ (where 
the equality sign applies only at the instants of time, 
$\tilde{t}_{k\in {\mathbb{Z}}}$, when the
 oscillators pass simultaneously through zero coordinate,
corresponding to the minimum position of the on-site potential, i.e. 
$q_{1,2}(\tilde{t}_k)=0$) can be
satisfied  and the proof is complete.

\hspace{16.5cm} $\square$

\vspace*{1.cm}

The next Theorem establishes the existence of localised out-of-phase solutions for motions in hard on-site potentials.

\vspace*{0.5cm}
\noindent {\bf Theorem 2:} 
{\it Let $(q_n(t),\dot{q}_n(t))$, $n=1,2$, be the smooth solutions 
to Eqs.\,(\ref{eq:start1}),(\ref{eq:start2}) with a hard on-site
potential and $-\infty<Q_l \le q_{1,2}(t) \le Q_r<\infty$ for $t \in {\mathbb{R}}$. 
There exist periodic solutions $(q_n(t+T_b),\dot{q}_n(t+T_b))=(q_n(t),\dot{q}_n(t))$ for  $n=1,2$,
so that   
the oscillators perform  out-of-phase motion, i.e.
 $\sign (q_1(t))=-\sign (q_{2}(t))$,
with period 
$T_b=2\pi/\omega_b$ and frequencies $\omega_b$ satisfying
\begin{equation}
 \sqrt{U^{\prime \prime}(0)}
 < \omega_b\le \sqrt{\max \{U^{\prime \prime}(Q_l),U^{\prime \prime}(Q_r)\}+2\kappa}\,. 
 \label{eq:ineq2}
\end{equation}
Moreover, these solutions are  localised  fulfilling either 
\begin{equation}
 |q_1(t)|> |q_{2}(t)|,\,\,\,t\in {\mathbb{R}},\,\,\,t \neq {\tilde{t}_k},\,\,\,k \in {\mathbb{Z}}
\end{equation}
or 
\begin{equation}
 |q_2(t)|> |q_{1}(t)|,\,\,\,t\in {\mathbb{R}},\,\,\,t 
 \neq {\tilde{t}_k},\,\,\,k \in {\mathbb{Z}}
\end{equation}
where $\tilde{t}_{k \in {\mathbb{Z}}}$ are the instants of time when the
 oscillators pass simultaneously through zero coordinate,
corresponding to the minimum position of the on-site potential,
i.e. 
$q_{1}(\tilde{t}_k)=q_{2}(\tilde{t}_k)=0$.
}

\vspace*{0.5cm}
\noindent {\bf Proof:} W.l.o.g. the initial conditions satisfy 
\begin{equation}
 q_{1}(0)=q_2(0) =0,\,\,\, 0<-\sign(\dot{q}_2(0))<\sign(\dot{q}_{1}(0)).
\end{equation} 
Then due to continuity there must exist some 
$t_*>0$ so that during the interval $(0,t_*]$ the two oscillators  perform 
motion with $\sign (q_1(t))=-\sign (q_{2}(t))$ and
$\sign(\dot{q}_1(t))=-\sign(\dot{q}_{2}(t))$.
Furthermore, on $(0,t_*]$ the 
following  order relation is satisfied
\begin{equation}
0 < - q_{2}(t) < q_{1}(t).\label{eq:order2}
\end{equation} 

We proceed as in the previous case for soft on-site potentials by introducing the difference variable between 
coordinates. The time evolution of the difference variable is determined by an equation identical to 
(\ref{eq:deltap}) and using the Lemma above 
we derive for $q_1(t)\ge -q_2(t)\ge 0$  for the r.h.s. an 
upper bound and lower bound for the r.h.s. for hard on-site potentials as
 \begin{equation}
\frac{d^2 \Delta q_*}{dt^2}\le -\omega^2_0\Delta q_*(t),
\end{equation}
 and 
\begin{equation}
 \frac{d^2 \Delta q_*}{dt^2}\ge  -(\Omega^2_h+2\kappa)\Delta q_*(t),
\end{equation}
respectively and $\Omega^2_h=\max \{U^{\prime \prime}(Q_l),U^{\prime \prime}(Q_r)\}$.

Thus, the solutions are bounded from above and below as $B(t)\le \Delta q_*(t)\le A(t)$ 
where the upper bound is given by 
\begin{equation}
   A(t)= \frac{\Delta \dot{q}_0(0)}{\sqrt{\Omega_h^2+2\kappa}}
   \sin(\sqrt{\Omega_h^2+2\kappa} \,t), \label{eq:qaboveh}
 \end{equation}
 and $0 < t <\pi/\sqrt{\Omega_h^2+2\kappa}$.
 \begin{equation}
   B(t)= \frac{\Delta \dot{q}_0(0)}{\omega_0}
   \sin(\omega_0\, t), \label{eq:qbelowh}
 \end{equation}
and $0 <t <\pi/\omega_0$.

The remainder of the proof regarding the time periodicity of 
the resulting localised solutions proceeds 
in an analogous way as above for Theorem 1.

Conclusively,  spatially localised and time-periodic solutions
for out-of-phase motion in hard on-site potentials result which satisfy
\begin{equation}
 |q_1(t)|\ge |q_{2}(t)|,\,\,\,t \in {\mathbb{R}},\label{eq:test}
\end{equation}
and $(q_n(t+T_b),\dot{q}_n(t+T_b))=(q_n(t),\dot{q}_n(t))$ for $n=1,2$ 
where 
the equality sign in (\ref{eq:test}) is valid only at the instants of time, $\tilde{t}_{k\in {\mathbb{Z}}}$, when the
 oscillators pass simultaneously through zero coordinate,
corresponding to the minimum position of the on-site potential, i.e. 
$q_{n}(\tilde{t}_k)=0$ with $-N \le n \le N$.  The period
$T_b=2\pi /\omega_b$ is associated with frequencies lying in the interval
\begin{equation}
 \sqrt{U^{\prime \prime}(0)} < \omega_b \le 
 \sqrt{\max \{U^{\prime \prime}(Q_l),U^{\prime \prime}(Q_r)\}+2\kappa}.
\end{equation}
The initial conditions can be chosen so that the
resulting amplitude of oscillations is such that the non-resonance 
condition between the amplitude-dependent breather frequency and the harmonic frequency,
$m \omega_b \ne \omega_0$ for all $m \in \mathbb{Z}$ is satisfied.

Furthermore, the considerations above imply that there exist initial conditions 
$-\infty <Q_l\le q_{1,2}(0)\le Q_r<\infty$
and $q_{1,2}(0) \ne 0$, $\dot{q}_{1,2}(0)=0$ resulting in 
time-reversible solutions obeying  $q_{1,2}(t)=q_{1,2}(-t)$ and 
$-\dot{q}_{1,2}(t)=\dot{q}_{1,2}(-t)$
 for $t \in {\mathbb{R}}$ completing the proof.

\hspace{16.5cm} $\square$

The in-phase and out-of-phase localised periodic solutions reduce to the harmonic linear normal modes at small amplitudes.

\vspace*{1.0cm}
Having established the existence of localised solutions for two coupled 
nonlinear oscillators we prove now the existence of time-periodic solutions exhibiting 
 patterns of two different amplitudes 
for a finite system of globally coupled nonlinear oscillators. 

\section{Breathing pattern solutions for globally coupled finite-size nonlinear lattices}

In the following we prove the existence of solutions to 
Eq.\,(\ref{eq:start}) in the form of breathing patterns where all sites have 
equal (comparatively small) 
amplitude (state $\{s\}$) except one with a large amplitude (state $\{L\}$). 
For the discrete 
nonlinear Schr\"odinger (DNLS) equation (arising  as a limit case of the coupled-oscillator lattice
considered in the current study)
the general exact expression for the corresponding
time-periodic solution that has equal amplitude on all sites except one  was given in \cite{DST}.

The existence of a single-site "breather" on a constant-amplitude tail can be anticipated 
from the work in \cite{Flach98} where it was found that for the general case 
of breathers on lattices
with long range interactions, which decay according to a power-law, 
breathers decay with the same power-law as the interaction. 
Hence, extrapolating this result to the case of equal interactions
(exponent of power-law going to zero), one can expect to get a constant-amplitude tail.

\vspace*{0.5cm}
\noindent {\bf Theorem 3:} 
{\it Let $(q_n(t),\dot{q}_n(t))$ be the smooth solutions 
to Eq.\,(\ref{eq:start})  with a soft on-site potential fulfilling 
$q_{i,-}<q_{l}\le q_n(t) \le  q_r<q_{i,+}$ for $-N\le n\le N$ and $t \in {\mathbb{R}}$. 
There exist periodic in-phase motions   
$(q_n(t+T_b),\dot{q}_n(t+T_b))=(q_n(t),\dot{q}_n(t))$ and 
$\sign (q_n(t))=\sign (q_{n +1}(t))$, 
with $T_b=2\pi/\omega_b$ and frequencies 
$\omega_b$ lying in the range 
\begin{equation}
\sqrt{\min\{U^{\prime \prime}(q_l),U^{\prime \prime}(q_r)\}}\le \omega_b < \sqrt{U^{\prime \prime}(0)+\kappa}.
\end{equation}
Moreover, these solutions are  characterised by a large amplitude at a  
single site of index $m$, compared to the equal small amplitude at 
the remaining sites, $n \neq m$, satisfying
\begin{equation}
 |q_m(t)| >|q_{-N\le n \le N,\,n\ne m}(t)|,\,\,\,t\in \mathbb{R},\,\,\,t 
 \neq {\tilde{t}_k},\,\,\,k \in {\mathbb{Z}},
\end{equation}
where $\tilde{t}_{k \in {\mathbb{Z}}}$ are the instants of time when all the
 oscillators pass simultaneously through zero coordinate,
corresponding to the minimum position of the on-site potential,
i.e. 
$q_{-N\le n \le N}(\tilde{t}_k)=0$.
}

\vspace*{0.5cm}
\noindent {\bf Proof:} W.l.o.g. the initial conditions are chosen such that 
$q_{n}(0)=0$ and 
$\dot{q}_{n}(0) \ne 0$,  say $\dot{q}_{n}(0) > 0$  
and $\dot{q}_{0}(0)>\dot{q}_{n \ne 0}(0)>0$  and equal 
$\dot{q}_{-N\le n\le N,\,n\neq 0}(0)$.
Then due to continuity there must exist some 
$\delta>0$ so that during the interval $(0,\delta]$ the 
following  order relations are satisfied
\begin{equation}
 0 < q_{-N\le n \le N,\,n\ne 0}(t) < q_{0}(t).
\end{equation}

We define the difference variable between the coordinates and velocities 
respectively of any pair of sites $k,l\in [-N,N],\,k,l\neq 0,\,l>k$ and 
$k=0,\,l\in [-N,N],\,l \neq 0$ as follows
\begin{equation}
\Delta q_{kl}(t)=  q_{k}(t)-  q_{l}(t),\,\,\,\Delta \dot{q}_{kl}(t)= 
\dot{q}_{k}(t)-  \dot{q}_{l}(t).
\end{equation}

The time evolution of the difference variables  $\Delta q_{kl} (t)$ is then 
determined by the following equation 
\begin{equation}
 \frac{d^2 \Delta q_{kl}}{dt^2}= -\left[U^{\prime}(q_{k})-U^{\prime}(q_{l})\right]
 -{\kappa} \Delta q_{kl}\label{eq:deltap1},
\end{equation}
which is closed in the variables $q_k$ and $q_l$, viz. is decoupled from the remainder of the lattice,
and represents the equation for the difference variable of two coupled oscillators (cf. Eq.\,(\ref{eq:deltap})).

If $k\ne 0$ it follows immediately from the assumptions on the initial conditions 
that $\Delta q_{kl}(t) \equiv 0$ and $\Delta \dot{q}_{kl}(t) \equiv 0$ for $t \in {\mathbb{R}}$. 
That is, $q_{k \neq 0}(t)=q_{l \neq 0}(t)$ and $\dot{q}_{k \neq 0}(t)=\dot{q}_{l \neq 0}(t)$. 
Furthermore, if $k=0$ we exploit the fact that 
for each pair of oscillators at  sites $k=0$ and $l\neq 0$ 
 there exists a localised and time-periodic 
solution as proven  
in Section \ref{section:two} such that $|q_0(t)|\ge |q_{-N\le n \le N,\,n\neq 0}(t)|$ and 
$q_{-N\le n \le N}(t+T_b)=q_{-N\le n \le N}(t)$ for $t\in {\mathbb{R}}$, where  the equality sign is valid only at the instants of time, $\tilde{t}_{k\in {\mathbb{Z}}}$, when the
 oscillators pass simultaneously through zero coordinate,
corresponding to the minimum position of the on-site potential, i.e. 
$q_{-N \le n\le N}(\tilde{t}_k)=0$.
Hence, a breathing pattern results for which all but one site 
attain the small-amplitude state $\{s\}$ while the remaining site is in the 
large-amplitude state $\{L\}$.

The frequency of oscillations satisfies the relations
\begin{equation}
\Omega_s\le \omega_b < \sqrt{\omega_0^2+\kappa}.\label{eq:ranges}
\end{equation}
and lies below the lower value of the linear spectrum.

In addition the  
non-resonance conditions $m\,\omega_b \ne \omega_{ph}$, $\forall m \in {\mathbb{Z}}$ have to be fulfilled, 
avoiding resonances of harmonics of the breathing pattern solutions 
with the linear (phonon) spectrum, which given that the latter comprises 
only two values is readily satisfiable by an appropriate choice of the initial conditions.

Furthermore, there exist initial conditions 
$q_{i,-}<q_{n}(0)<q_{i,+}$ with $|q_0(0)|>|q_{n \neq 0}(0)|$ and $\dot{q}_n(0)=0$ 
with $-N \le n \le N$, 
so that time-reversible solutions $q_{n}(t)=q_{n}(-t)$ 
and $-\dot{q}_{n}(t)=\dot{q}_{n}(-t)$ with $q_{i,-} <q_n(t)<q_{i,+}$ for $-N \le n \le N$ 
$|q_{n\neq 0}(t)|\le |q_0(t)|$ and 
$t \in {\mathbb{R}}$  where 
the equal sign applies only at the instants of time, $\tilde{t}_{k\in {\mathbb{Z}}}$, when the
 oscillators pass simultaneously through zero coordinate,
corresponding to the minimum position of the on-site potential, i.e. 
$q_{-N \le n\ le N}(\tilde{t}_k)=0$, result completing the proof.

\hspace{16.5cm} $\square$

\vspace*{1.0cm}
Regarding the existence of solutions to 
Eq.\,(\ref{eq:start}) in the form of breathing out-of-phase patterns  where all the oscillators 
possess equal small amplitude except for one oscillator evolving with large amplitude
in hard on-site potentials we have the following theorem which 
is proven in analogous manner to the previous proof:

\vspace*{0.5cm}
\noindent {\bf Theorem 4:} 
{\it Let $(q_n(t),\dot{q}_n(t))$ be the smooth solutions 
to Eq.\,(\ref{eq:start})  with a hard on-site potential and 
$-\infty<Q_l \le q_{n}(t) \le Q_r<\infty$ for $0\le n\le N$  and $t \in {\mathbb{R}}$.  
There exist periodic out-of-phase motions   
$(q_n(t+T_b),\dot{q}_n(t+T_b))=(q_n(t),\dot{q}_n(t))$ and 
$\sign (q_n(t))=-\sign (q_{n +1}(t))$,  
with $T_b=2\pi/\omega_b$ and frequencies 
$\omega_b$ lie in the range 
\begin{equation}
\sqrt{U^{\prime \prime}(0)}< \omega_b\le 
 \sqrt{\max \{U^{\prime \prime}(Q_l),U^{\prime \prime}(Q_r)\}+\kappa}. 
\end{equation}
Moreover, these solutions are  characterised by a large amplitude at the  
single site, w.l.o.g. chosen as $n=0$, compared to the equal amplitude at the remaining sites, $n \neq 0$,
satisfying
\begin{equation}
 |q_0(t)| >|q_{-N\le n \le N,\,n\ne 0}(t)|,\,\,\,t\in \mathbb{R},\,\,\,t 
 \neq {\tilde{t}_k},\,\,\,k \in {\mathbb{Z}}
\end{equation}
where $\tilde{t}_{k \in {\mathbb{Z}}}$ are the instants of time when all the
 oscillators pass simultaneously through zero coordinate,
corresponding to the minimum position of the on-site potential,
i.e. 
$q_{-N\le n \le N}(\tilde{t}_k)=0$.
}

\vspace*{1.0cm}
{\it Remark:} Theorems $3$ and $4$ apply also to lattices with an even number of sites, 
$-N+1 \le n \le N$ (or $-N \le n \le N+1$).

\vspace*{1.0cm}
\noindent {\bf Corollary:} With the methods introduced in this manuscript 
one  also shows the existence of 
{\it multi-site breathing pattern solutions}, comprising any combination of the two
states $\{s\}$ and $\{L\}$. 
 To this end one defines
 the distance variables $\Delta q_{kl} (t)$ between pairs of oscillators at 
  sites $k$ and $l$ as $\Delta q_{kl}(t)=\alpha[q_k(t)-q_{l}(t)]$,
 when $|q_k(t)|\ge |q_{l}(t)|$ with $\alpha=\sign(q_k)$, and 
 $\Delta q_{kl}(t)=\alpha[q_{l}(t)-q_{k}(t)]$,
when $|q_{l}(t)|\ge |q_{k}(t)|$ with $\alpha=\sign(q_{l})$. 
Unless the oscillators at sites $k$ and $l$ are supposed to perform 
motions with equal amplitude (i.e. they are both either
in state $\{s\}$ or in state $\{L\}$) 
one  facilitates the 
 methods contained 
 in the proofs above to establish the existence of a localised state at sites $k$ and $l$ where the 
 site $k$ is 
 in state $\{s\}$ while site $l$ is in state $\{L\}$ or vice versa.

\vspace*{1.0cm}

\section{Summary}

We have proven the existence of time-periodic solutions comprising two states of 
different amplitudes 
in finite nonlinear lattices whose units are globally coupled, utilising the 
comparison principle for differential equations. It is demonstrated 
that spatial patterns are possible, that  
are built up from any combination of 
a small-amplitude state $\{s\}$ and a large-amplitude state $\{L\}$. 
The spectrum of linear solutions (phonons) consists of two values only. 
For soft (hard) on-site potentials the in-phase (out-of-phase) breathing patterns
 are allowed to have frequencies  below (above) the lower (upper) value of the 
 bivalued degenerate linear spectrum of phonon frequencies.
Due to the very discrete nature  of the linear spectrum,  
resonances between harmonics
of the breathing solution frequency and the 
frequencies of linear oscillations, i.e. phonons, 
are readily avoidable.

\end{document}